# Empower Children in Nigeria to Design the Future of Artificial Intelligence (AI) through Writing

Cornelius Adejoro

University of Colorado Boulder, cornelius.adejoro@colorado.edu

Luise Arn

University of Zurich, arn@ifi.uzh.ch

Larissa Schwartz

University of Colorado Boulder, larissa.schwartz@colorado.edu

Tom Yeh

University of Colorado Boulder, tom.yeh@colorado.edu

**ABSTRACT**

This paper presents a new approach to engaging children in Nigeria to share their views of AI. This approach is centered on an inclusive writing contest for children in a secondary school in Abuja to write about AI to compete for prizes and share their writings with others. A preliminary analysis of the first 11 articles we received exhibits diverse gender and ethnic representation that conveys cultural values and perspectives distinct from those of the children in the western countries. This finding suggests future work to conduct in-depth cross-cultural analysis of the articles and to replicate similar writing contests to engage children in other underrepresented countries.

## 1. INTRODUCTION

"Ina bukatar basirar wucin gadi ya taimakawa yan uwa na maza da mata." [1]

"Gbogbo ènìyàn ló yè kí ó kó nípa òye atọwọ́dá." [2]

"Kedụ ka amamịghe echiche ga-esi nyere ezinaụlọ m aka?" [3]

Children in Nigeria, such as those who speak Hausa, Yoruba, and Igbo, have ideas, views, dreams, and concerns about Artificial Intelligence (AI) and its impact on their future. Their voices, however, are seldom heard. More broadly speaking, children in Sub-Saharan Africa (SSA), where Nigeria is the most populous country, are underrepresented in the human-computer interaction (HCI) literature. There have been limited opportunities for SSA children to participate in the design process of AI-based solutions. Without being involved and providing input, SSA children are unlikely to enjoy the same level of benefits compared to their counterparts in Western countries. To address this gap, we developed a novel *Inclusive Writing Contest* as a method to empower SSA children to participate in the design of the future of AI. This paper reports the work in progress in our effort to test and study the effectiveness of this method in Nigeria, with the hope that our findings can generalize to other SSA countries in the future.

---

[1] In Hausa. English translation: "I want Artificial Intelligence to help my brothers and sisters."

[2] In Yoruba. English translation: "Everyone should learn about Artificial Intelligence."

[3] In Igbo. English translation: "How can Artificial Intelligence help my family?"

## 2. BACKGROUND

A study [1] compared children between Uganda and Japan about their views of what a "good" robot would do in certain scenarios, such as serving as a gatekeeper to decide how to treat a child running late to school---let in, penalize, or deny entry. Significant cross-cultural differences were observed in children's views, with important implications on the design for social robots. This finding shows the value of including underrepresented children from other cultures in HCI research. Instead of focusing on specific scenarios, our writing contest keeps it open-ended by inviting children to write anything on their mind about AI.

Prior to our effort in Nigeria, we conducted a small pilot in Brazil, which drew six submissions from a single classroom in a school in Fortaleza, a northern coastal city with a diverse population (63.2% Pardo, Black, Asian). Here is an excerpt from one of the winners:

> (Original in Portuguese) Chegando no setor ético e visando se os direitos humanos seriam aplicados a esta inteligência virtual, já que por um lado ele é autoconsciente possuindo mente própria e uma opinião formada como um ser pensante, por outro lado, ele é fruto de um código de programação que poderia ser reproduzido em massa. Assim abrimos espaço para indagarmos, caso seu código seja reproduzido em massa, ele ainda possuiria uma identidade predominantemente única ou seria educado com base no controle de informações expostas ao novo indivíduo criado a partir de uma série de códigos de programação, como se fosse uma criança?

> (Translation in English) Ethically speaking, we must wonder if human rights would be applied to this virtual intelligence since on the one hand it is self-aware, possessing its own mind and an opinion formed as a thinking being, but on the other hand, it is the fruit of a computer code that could be mass reproduced. So, we should ask if the code is mass reproduced, would it still possess a predominantly unique identity or would it be educated based on the control of the information it is exposed to which in turn is created from a series of other codes as if it were a child?

The writings of Brazilian children, albeit a small sample, demonstrated the potential of an inclusive writing contest as a method to engage children outside of western countries to share their views about AI.

Research practices need to be such that they include more diverse voices in design processes. Namely, they should be less geographically bound to the researchers [6]. With this in mind, we solicit voices on AI-based solutions from previously excluded areas, such as Nigeria. Nigeria has the highest level of AI innovation, among the Sub-Saharan African (SSA) countries, with 42 AI use cases found in the sample-based case study [3]. While some AI systems are starting to emerge into everyday life in Nigeria, such as financial and agricultural service technologies, other AI systems have been unable to acquire attention due to missing infrastructure or legal, ethical, or health concerns [4]. With efforts such as Artificial Intelligence for Development Africa and EduAI Hub, Nigeria as a country, and Africa as a whole, work towards the goal for responsible development and use of AI [3]. Nevertheless, despite their history of substantial technological development, Nigerian voices on AI are rarely included in international scientific publications or global expert committees [2, 5]]. To find effective and sustainable global solutions and policies for a world with AI, different perspectives and contexts need to be considered by the AI community [5]. For this work, we can build on [1]], who used storytelling activities in an educational setting to solicit children's reflections on AI fairness for a cross-cultural comparison between Japanese and Ugandan children. We suggest that the short essays of Nigerian students on AI-based solutions are one way to make Nigeria's AI voices heard.

## 3. METHOD

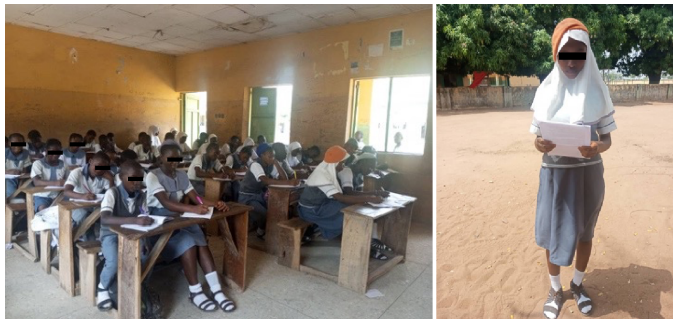

The main goal of our study is to empower diverse children in Nigeria to design the future of AI by providing a platform for them to express and share their views. Our method is to organize an *Inclusive Writing Contest*.

**Location:** We partnered with a secondary school in Abuja, Nigeria, which is the administrative and political capital of the country and the most diverse city in terms of ethnicity and religion (i.e., even representation of Muslim and Christian children in schools).

**Children:** The writing contest for Nigerian children was held in a secondary school. A total of thirty students between the ages of 10–13 signed up to participate in the writing contest. Of the participants, 14 were girls and 16 were boys. We also wanted the culture and background of participants to reflect in the contributions, therefore we drafted participants from across different ethnic groups that spoke different languages, which included Hausa, Yoruba, and Igbo.

**Teachers:** We adopted the participatory design approach by partnering with teachers in Abuja in the planning, execution, and judging of the contest. Our partner teachers were recruited from two different schools. While some are from the school where the writing contest took place, we deliberately decided to involve teachers from another school to work with them to minimize the distortion of the results or output of the students.

**Writing Contest:** By presenting our project as a *Writing Contest* as opposed to a regular programming contest or a hackathon event, we aimed to reduce the barrier for participation. School children do not need to know how to program, nor do they need to have access to a powerful computer and a fast Internet connection; they just need to be willing to share their thoughts, feelings, opinions, and views about AI and write. Below is a summary of the design of our writing contest:

- *Classroom Learning:* Prior to the contest, our partner teachers gave short lessons about AI to students in their class and encouraged them to participate in the AI writing contest. These lessons are meant to prepare students for writing, and not to control their writing in any way; students were able to freely express themselves.
- *Short Writings*: Students were encouraged to write short essays of 250-500 words.
- *Local Competition*: Students only compete with their classmates in the same school, not with other schools.

- *Any Medium*: Since all our participants could write by hand, only hand-written stories on paper were accepted. This was to ensure students without access to computers were given fair access to participate.
- *Local Language*: Participants were encouraged to write in their native language, but because Abuja is a cosmopolitan city where most students write in English; all our participants wrote their stories in English and hence we did not need experts to translate at this time.

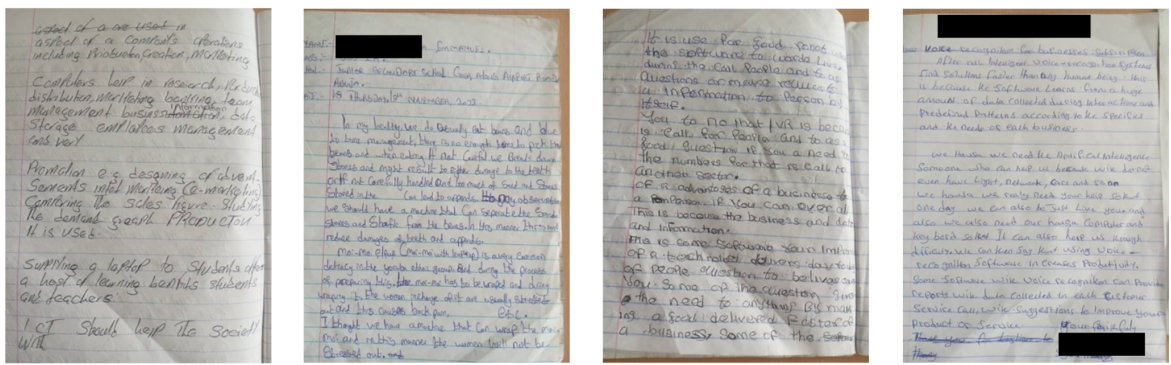

**Figure 2: Examples of Student's Hand-Written Essays about AI**

- *Motivation*: We incentivized participants by promoting our activities as AI Writing Contests that offer scholarships for top winners (i.e., $100, $50, $30 USD for 1st, 2nd, and 3rd place). For every participant, we issued certificates (signed by a university), as well as a small gift of appreciation (such as a USB flash drive).

## 4. RESULTS

At the time of repairing this work-in-progress report, we received a total of 11 submissions. *Table 1* provides a summary. *Figure 2* shows a sample of students' hand-written essays submitted to our writing contest. We included a few excerpts below. Sentences where children refer to themselves or people in their communities are underlined.

**Table 1: Submissions we received from Nigerian school children (work in progress)**

| ID | Gender | Tribe | Title | # of Words |
|---|---|---|---|---|
| 1 | M | Yoruba | IoT in agriculture – developing smart farming in Nigeria | 765 |
| 2 | F | Hausa | AI inventing indestructible cars | 177 |
| 3 | F | Igbo | Areas we need artificial intelligence in Africa | 761 |
| 4 | M | Hausa | How can artificial intelligence be useful to us in our country | 154 |
| 5 | F | Yoruba | Language translator | 261 |
| 6 | M | Igbo | - | 168 |
| 7 | M* | Yoruba | Making cars with the use of solar panels | 166 |
| 8 | F | Igbo | Voice recognition for businesses subscription | 100 |
| 9 | M | Yoruba | - | 154 |
| 10 | F | Igbo | How AI can be useful in our society, culture and country / continent at large | 242 |
| 11 | M | Igbo | Virus detector | 122 |

*"Search machine is a very brilliant tool used to search for important information on the internet as a means of information. With the help of voice recognition, our languages can be translated to us through the voice recognition system with intelligence such as the language translating system will help a lot of people here in Nigeria, because not all Nigerians are able to speak and write with English because they prefer their own language to English."*

*"Technology has been of great help to everybody and it also causes damage to we human beings. Technology is usually found in the city. But, there are places where technology is not usually seen. Like in my place Delta state, technology is not usually seen there, so if I am given the opportunity to construct any machine, the first technological machine I will construct is a farming machine."*

*"One of the benefits of using IoT in agriculture is the increased agility of the processes. Thanks to real-time monitoring and prediction systems, farmers can quickly respond to any significant change in weather, humidity, air quality as well as the health of each crop or soil in the field. In the conditions of extreme weather changes, new capabilities help agriculture professionals save the crops."*

*"AI is very important to everyone as it helps in new inventions. Digital assistance, reduction in human error brings about automation, it helps to solve complex problems, minimize errors and helps humans make smarter and faster decisions."*

*"With artificial intelligence, our work will be faster, stress-free, risk-free, and more efficient. As much as artificial intelligence is of great help in human endeavor and is greatly desired, the high cost of acquisition and maintenance and inadequate supply of electricity has made it unavailable in our community."*

*"We, the Hausas need artificial intelligence, someone who can help us because we do not even have lights, a network, cars, etc. We need a Hausa computer keyboard so that it can also help us through difficulty, so we can then say that using voice recognition software increases productivity."*

*"Why? This is because some Igbo people believe that if you have large farms, you should marry more than one wife and have many children that will help make work on the farm easy, but if the farm machines are available, it will reduce their belief of marrying many wives."*

## 5. ANALYSIS

We performed a preliminary cross-cultural comparison analysis. We recruited US American high school students to read through the writing samples. We recruited high school students through our lab's network. We instructed the two students to highlight statements in the writing samples that they agreed or disagreed with. This opening task was designed to ease the students into the examination of the international writing samples, while simultaneously providing us with a first intuition on feasible directions for our cross-cultural comparison analysis.

Regarding the Nigerian writing samples, the two US American students, for instance, agreed with AI being helpful in many ways and making life more comfortable, as well as AI already being applied in many areas, while there still exist many more promising application areas for AI. For instance, both students agreed with the following three statements: *"The following are some benefits of using IoT in agriculture: Improved efficiency [... ,] Effective use of resources [..., and] Clean and organic process.", "It [an AI system] saves time and energy […]."*, and *"It will be of great help and development to man."* The students' agreement with these statements could indicate a similar level of awareness and appreciation for AI across cultures. Both US

American high school students also had some disagreements. For example, both disagreed with the statement "*voice recognition provides a good experience to consumers*", as they felt "*customers may not like conversing with an automated voice*". Additionally, the US American high school students noticed a focus on AI in the automation of farming throughout the Nigerian writing samples.

Collectively, our preliminary data analysis indicates that there are many similarities across cultures when it comes to the perception of AI and its potential. There are also indications for culturally dependent AI implementations and potential implications, as well as some differences in perceptions. For more conclusive comparisons, we are currently conducting a more in-depth cross-cultural comparison analysis, where we include additional dimensions and the perspectives from further US American high school students.

## 6. CONCLUSION AND FUTURE WORK

We presented a work-in-progress study on the feasibility of an *Inclusive Writing Contest* as a method to enable Nigerian children to share their views about AI with the rest of the world. Our early results suggest that diverse gender and ethnic (Hausa, Igbo, Yoruba) representation of children can be achieved. The writings by those children reveal perspectives and values unique to their cultures. A cross-cultural analysis by US high-school students also revealed both similarities and differences. Our future work includes completing the ongoing writing contest in Nigeria (i.e., receiving more submissions), performing a comprehensive analysis of Nigerian children' writings, replicating similar inclusive writing contests to empower more children in other countries, and extracting new AI design guidelines that will reflect the voices of those empowered children.

**ACKNOWLEDGMENTS**

**REFERENCES**

[1] Vicky Charisi, Tomoko Imai, Tiija Rinta, Joy Nakhayenze, and Randy Gomez. 2021. Exploring the Concept of Fairness in Everyday, Imaginary and Robot Scenarios: A Cross-Cultural Study With Children in Japan and Uganda. In Interaction Design and Children (IDC '21), June 24–30, 2021, Athens, Greece. ACM, New York, NY, USA, 5 pages. https://doi.org/10.1145/3459990.3465184

[2] Xieling Chen, Haoran Xie, Di Zou, and Gwo-Jen Hwang. 2020. Application and theory gaps during the rise of artificial intelligence in education. Computers and Education: Artificial Intelligence 1: 100002.

[3] Arthur Gwagwa, Patti Kachidza, Kathleen Siminyu, and Matthew Smith. 2021. Responsible artificial intelligence in Sub-Saharan Africa: landscape and general state of play. URL: http://hdl.handle.net/10625/59997 (visited March-02-2023).

[4] Nugun P. Jellason, Elizabeth JZ Robinson, and Chukwuma C. Ogbaga. 2021. Agriculture 4.0: Is sub-saharan Africa ready?. Applied Sciences 11, no. 12: 5750.

[5] Anna Jobin, Marcello Ienca, and Effy Vayena. 2019. The global landscape of AI ethics guidelines. Nature Machine Intelligence 1, no. 9: 389-399. DOI: https://doi.org/10.1038/s42256-019-0088-2

[6] Greg Walsh. 2018. Towards equity and equality in American co-design: a case study. In Proceedings of the 17th ACM Conference on Interaction Design and Children (IDC '18). Association for Computing Machinery, New York, NY, USA, 434–440. https://doi.org/10.1145/3202185.3202768